# Optical probing of charge traps in organic field-effect transistors

Dean Kos* and Marta Mas-Torrent

Institut de Ciència de Materials de Barcelona (ICMAB-CSIC), Campus UAB, 08193 Bellaterra, Spain

* corresponding author email: deankos@icmab.es

## Abstract

**We report spatially resolved optical probing of charge traps in organic field-effect transistors using focussed laser illumination. By scanning a 635 nm laser across the transistor channel and simultaneously acquiring transfer characteristics, we observe persistent, localised shifts in transistor turn-on voltage correlated with illumination dose and position, with negligible impact on field-effect mobility. The effect is strongest 5–10 μm from the source electrode and requires a drain-to-source scan direction with sub-10 μm step size. Kelvin probe force microscopy confirms trapped negative charges along the scan path, consistent with exciton dissociation and electron trapping near the semiconductor–dielectric interface. The phenomenon is reproducible across multiple device geometries and organic semiconductors, including TMTES-pentacene, TIPS-pentacene, and diF-TES-ADT. These findings enable direct mapping of trap distributions and suggest new strategies for trap engineering, threshold voltage tuning, and development of organic optoelectronic memories.**

## Introduction

Charge traps in organic semiconductors are spatially localised charge carrier states with energy within the semiconductor band gap, and they have been attracting considerable interest for their role in charge transport.[1] In many cases their impact on transport is undesirable, reducing carrier mobility, inducing hysteresis in electrical characteristics, slowing down electrical response, shifting the threshold voltage in transistors, and assisting non-radiative recombination in emissive devices.[2–4] In other cases, such as in sensors, traps are necessary because they are responsible for the sensing response, and are therefore a key element in chemical sensors, temperature sensors, and photodetectors, among others.[5–8] Whether charge traps are beneficial or not, their understanding and rational engineering are essential to achieve high-performance devices. A challenge in investigating charge traps is that often their characteristics can only be evaluated indirectly from macroscopic device response, when they are actually local in nature and usually occur at material interfaces, grain boundaries, doping sites, and other confined defects.[9,10] Traps of different types in different locations combine with primary device properties to yield a macroscopic response where individual contributions are difficult to disentangle. Organic field-effect transistors (OFETs) have proven a very useful platform in this context, as the gate voltage provides an additional degree of freedom to interact with traps at the semiconductor-dielectric interface, which is where charge transport occurs.

Here we measure an OFET while its channel area is simultaneously scanned by a focussed laser beam with energy above the semiconductor band gap. Photoexcited electron-hole pairs (excitons) interact with charge traps and modify the device response, which is continuously tracked at each illumination point. This allows us to pinpoint any changes in transistor transfer characteristic to a specific location, time, and illumination dose. Some previous works have employed light to locally probe an OFET channel. The transfer curves of OFETs made with acene-based molecules, diF-TES-ADT, and C8-BTBT have been shown to reversibly shift upon exposure to various types of illumination, and the phenomenon is widely attributed to charge trapping.[11–15] In some cases this shift is accompanied by a photoconductive effect where the drain-source current ($I_{DS}$) increases in the transistor off state.[16] The shift primarily concerns the threshold voltage ($V_{th}$), and has been likened to a bias stress effect accelerated by light.[17] In one report, $I_{DS}$ was monitored under focussed illumination with fixed gate- ($V_{GS}$) and drain-source ($V_{DS}$) voltages, observing current fluctuations which revealed that electrode interfaces can be hotspots for charge traps.[18] In another work, light was reported to neutralise carrier traps with concurrent release of majority carriers in the channel, contributing to the overall $I_{DS}$.[19] The central role of charge traps in photodetection is also recognised by a large body of literature on organic phototransistors, but these works are generally more focussed on improving the performance of devices as detectors rather than unearthing the basic mechanisms of light interaction.[20–23]

Changes in OFET transfer characteristic and key device parameters such as mobility and threshold voltage were so far never investigated systematically as a function of focussed illumination with high spatial resolution.

## Results

OFETs are fabricated on a doped Si/SiO$_2$ substrate in a bottom gate, bottom contact geometry (see Methods). The bottom Si of the substrate is used as gate, whereas the source and drain electrodes are patterned by photolithography to produce a transistor channel of width W=100 µm and length L=40 µm, unless otherwise noted. The electrode pattern is coated with a blend of 1,4,8,11-tetramethyl-6,13-triethylsilylethynyl pentacene (TMTES-pentacene) and polystyrene (PS) by solution shearing to form a uniform polycrystalline film (see Methods). PS is used to improve processability and is known to segregate into a separate phase during deposition, forming an additional dielectric layer underneath TMTES-pentacene that improves both crystal order and electrical performance.[24–26] As observed by the optical polarised microscopy images, away from the electrodes the polycrystalline TMTES-pentacene domains are randomly shaped with typical size of 50-100 µm, while on top of the electrodes grains are much smaller, mostly <10 µm in diameter (**Figure 1a**). Electrodes serve as nucleation points for crystal growth, so along electrode edges grains are narrow and elongated perpendicularly to the edge. In the transistor channel this leads to crystal domains of approximately equal length extending from the electrodes, yielding a continuous central grain boundary running roughly parallel to the electrodes along the middle of the channel. TMTES-pentacene is a p-type semiconductor and we obtain the typical transfer characteristics expected for this type of material (**Figure 1b**). $V_{th}$ for newly fabricated devices is typically in the 0-2 V range, with a sharp turn-on at gate-source voltage ($V_{GS}$) lower than $V_{th}$ and no hysteresis in the transfer response. The drain-source current ($I_{DS}$) in the off state ($V_{GS}>V_{th}$) is <200 pA and comparable to the noise floor of our equipment at the used settings, as is the gate-source leakage current ($I_{GS}$) that remains indistinguishable from noise for all measurements reported here. The field-effect mobility ($\mu_{FE}$) in saturation regime is generally between 2 and 3 $cm^2V^{-1}s^{-1}$, independent of $V_{GS}$, with a maximum current around 1 µA at $V_{GS}$=-5 V and $V_{DS}$=-10 V.

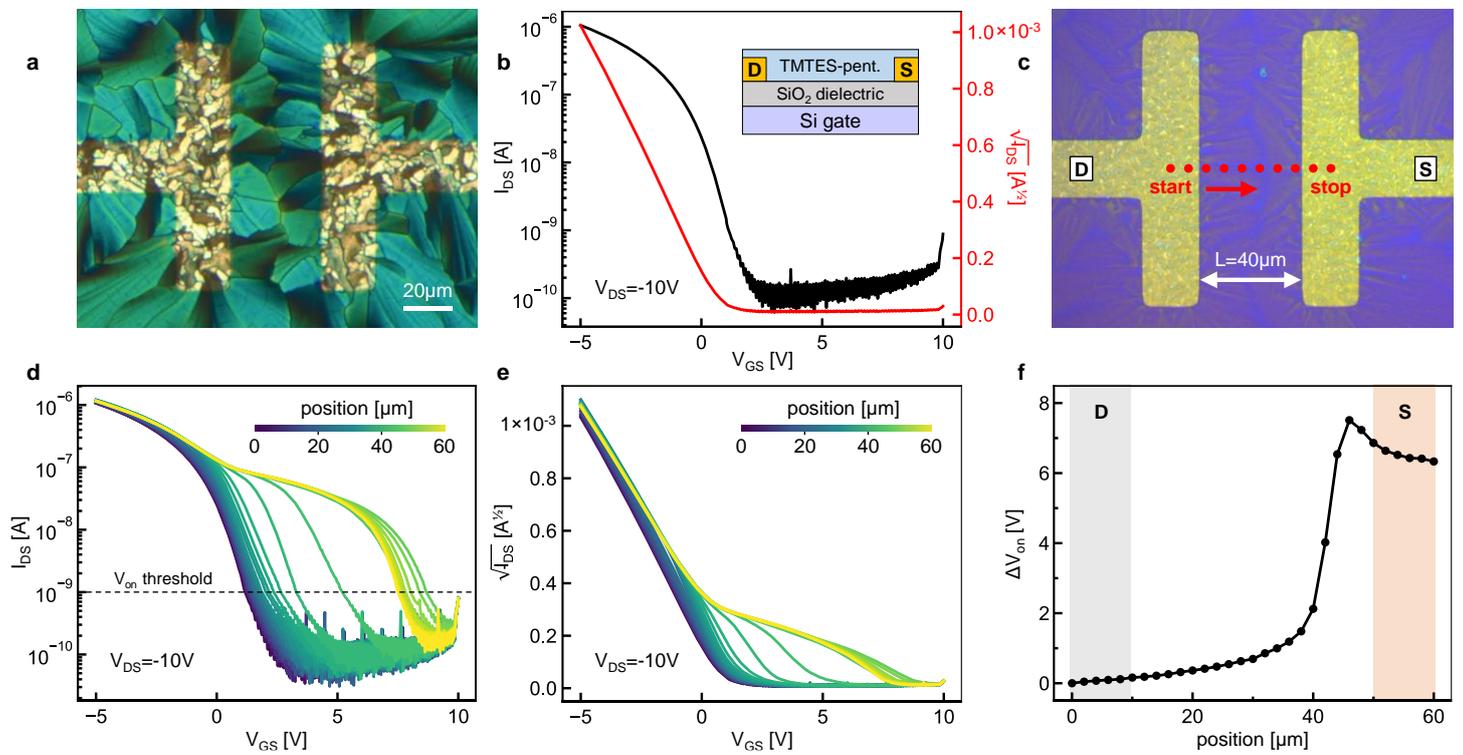

**Figure 1. OFET photoresponse**. **a**, Polarised optical microscopy image of the OFET device, with elongated crystals forming a grain boundary along the middle of the channel. **b**, Channel and contact geometry, and representative transfer characteristic. **c**, Typical scanning experiment where laser spot moves from drain to source illuminating discrete points along the path (size and number of spots not to scale). **d**, Transfer curves at each position along the scan in c, in log scale and **e**, as $\sqrt{I_{DS}}$. Dashed line in d indicates threshold for $V_{on}$ calculation. **f**, Shift in $V_{on}$ along the scan, referenced to $V_{on}$ at the start of the scan.

Optical probing of the OFET is done with a 635 nm laser focussed onto a circular diffraction-limited spot with diameter of 1 µm whose position on the sample is controlled with a motorised stage. At the start of an experiment, a grid of discrete points to be probed is defined on the OFET area (**Figure 1c**), along with the $V_{DS}$ value and $V_{GS}$ range to be used for acquisition of transfer curves. The system is positioned on the first point of the grid and a transfer curve is measured in the dark (laser turned off). Then, the laser is turned on and another transfer curve is measured with the same settings maintaining illumination throughout the measurement. The system then moves to each subsequent grid point, repeating the procedure until the scan is completed. Figure 1c shows a representative linear scan, where the laser scans from drain to source in 2 µm steps with an optical power of 1 µW, covering 10 µm of each electrode and the full 40 µm length of the channel. Transfer curves under illumination (**Figure 1d**) remain largely the same until about 40 µm into the scan, meaning 30 µm into the channel from the drain side, when the transistor turn-on voltage abruptly shifts to more positive values. The biggest changes are localised to just a few grid points in this region <5 µm apart overall, then saturate and marginally recover as the scan progresses into the source electrode. At the grid points where photoresponse is observed, $I_{DS}$ sees the largest changes for $V_{GS}>V_{th}$ (the original $V_{th}$ of the transistor) going from <1 nA to >50 nA, whereas for $V_{GS}<0$ V where $I_{DS}$ was already large the relative change is small. This implies that $\mu_{FE}$, as extracted for the $V_{GS}<0$ V range, remains mostly unaffected (**Figure 1e**). To characterise the observed response and avoid confusion with the established notion of $V_{th}$, we define the parameter $V_{on}$ as the $V_{GS}$ value at which $I_{DS}$ reaches a threshold that is unequivocally above the noise. This threshold is set to $I_{DS}=1$ nA unless stated otherwise (dashed line Figure 1d). $V_{on}$ simply quantifies the change in transfer without fitting the data with a model. To account for slight natural variations in $V_{on}$ at the start of a scan between different devices (or scans, if the same device is scanned multiple times), we further calculate $\Delta V_{on}=V_{on}-V_{on,start}$ where $V_{on,start}$ is the $V_{on}$ value of the first point of the scan. Plotting $\Delta V_{on}$ vs position along the scan (**Figure 1f**) thus quantitatively evaluates the change in transfer response as a result of illumination, and highlights the sharp shift in $V_{on}$ 5-10 μm away from the source.

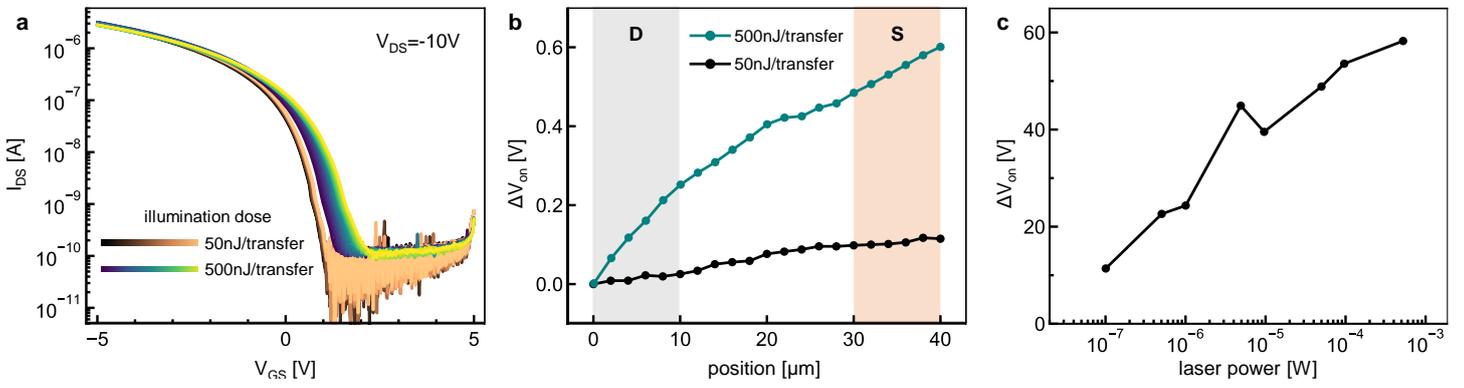

**Figure 2. Illumination dose dependence of photoresponse. a**, Transfer curves remain unchanged for illumination dose of 50 nJ per transfer, but shift for 500 nJ per transfer. **b**, Quantitative comparison of shift in a. **c**, $\Delta V_{on}$ increases with optical power throughout the measured power range.

The positive shift in $V_{on}$ is observed systematically in all measured samples (>200 devices in total), with qualitatively similar displacement of the transfer curve. When left in the dark following a completed linear scan, the transfer characteristic gradually shifts back by 10-20% of the maximum $\Delta V_{on}$ until settling on a stable response thereafter. The change induced by localised illumination is therefore permanent, so devices are never reused for multiple experiments unless explicitly noted. The minimum laser power required to stimulate a response was investigated by fabricating a device with shorter a channel length of L=20 μm, thus reducing the total time required for a linear scan, and dropping laser power to 5 nW. Considering the measurement time of a transfer curve, this corresponds to an illumination dose of 50 nJ per scan point. No shift in $V_{on}$ was observed at this dose level (**Figure 2a**) with a $\Delta V_{on,max}=0.1$ V over the entire scan (**Figure 2b**), compatible with the drift of transfer curves measured repeatedly in the dark. The scan was immediately repeated on the same device over exactly the same scan points, but adding a short delay in data acquisition to achieve a dose of 500 nJ per transfer curve. With this higher dose a clear $\Delta V_{on}$ is observed (Figure 2a,b), with $\Delta V_{on,max}=0.6$ V at the end of the scan. This result indicates that a dose of 500 nJ per scan point is the minimum to detect a photoresponse in these conditions, and that delivered illumination dose is indeed the quantity that determines the response, rather than optical power. We note that the gradual increase in $\Delta V_{on}$ of Figure 2b, as opposed to the sharp localised response of Figure 1f, is typical of these very low doses of <1 μJ/transfer. Exploring the behaviour of $\Delta V_{on}$ at very high dose levels is not straightforward, since adding further delays in data acquisition rapidly escalates the time for a single scan to several hours. This makes measurement impractical and causes prolonged electrical stress on the device, so increasing optical power is preferable to increase dose with the same illumination time. As $\Delta V_{on}$ increases with dose however, the $V_{GS}$ range needs to be progressively widened to accommodate the increasing $V_{on}$, so the $V_{GS}$ step size and range have to be balanced to maintain a constant illumination time.

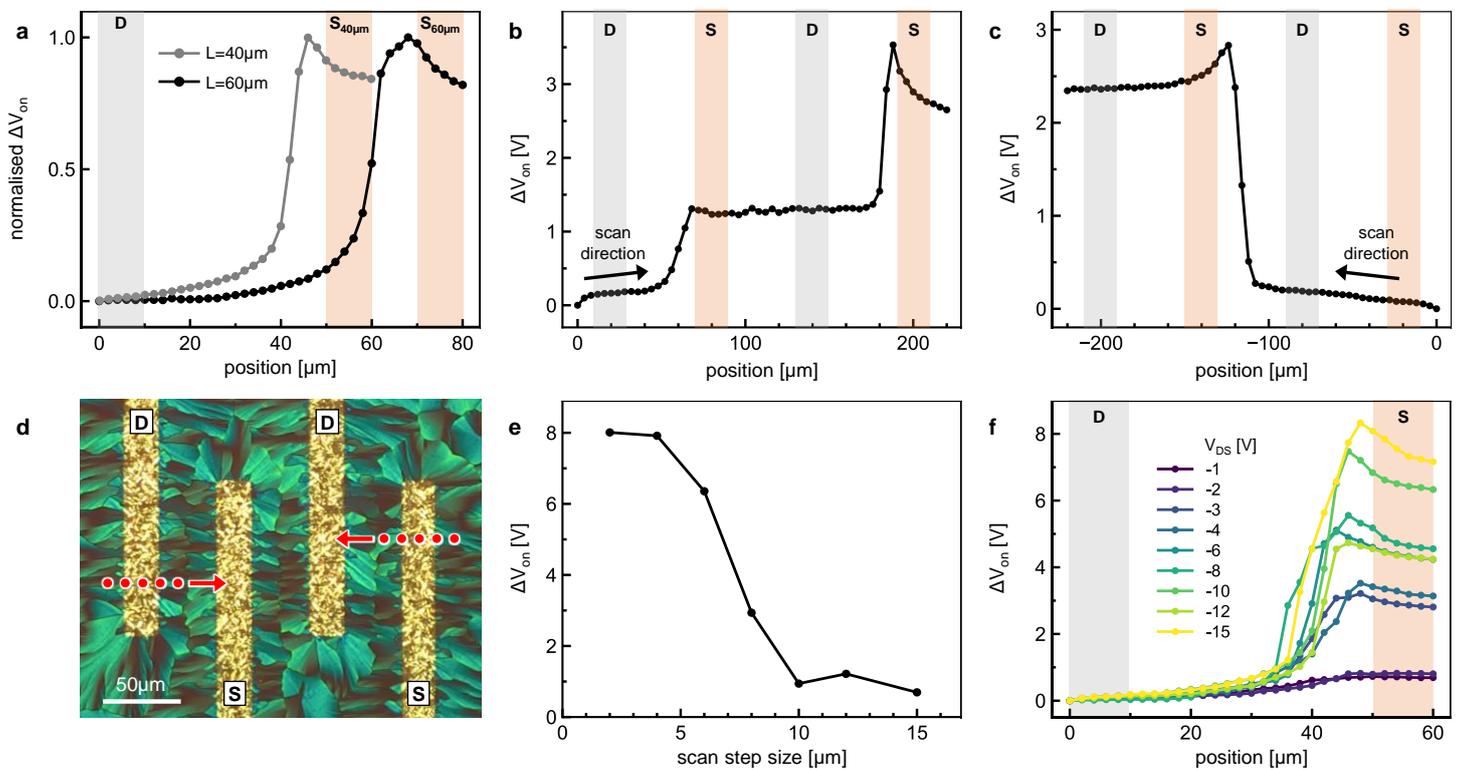

**Figure 3. Impact of optical probing location and $V_{DS}$ on transfer shift. a**, $\Delta V_{on}$ normalised to $\Delta V_{on,max}$ vs scan position for L=40 μm (grey) and L=60 μm (black) devices. The biggest shift in $V_{on}$ occurs near the source electrode regardless of channel length. **b**, $\Delta V_{on}$ along a forward and **c**, backward scan of the interdigitated electrodes. **d**, Polarised microscopy image of interdigitated electrode OFET with marked forward and reverse scan directions. **e**, $\Delta V_{on}$ vs the scan step size, showing that it drops off quickly as the step size is increased above 6μm. **f**, $\Delta V_{on}$ as a function of $V_{DS}$.

**Figure 2c** summarises the results on increasing illumination power for a set of 40 μm channel devices, where illumination time is constant and equal for all points and thus dose scales directly with power. $\Delta V_{on}$ clearly increases with laser power, reaching values as high as 60 V and appearing to saturate above 0.1 mW. Such level of transfer curve displacement is striking, considering that it originates from a set of illumination spots each 1 μm in diameter, that all together cover 0.5% of the channel area.

The change in transfer response over an illumination scan is similar in devices with different channel lengths. **Figure 3a** compares the normalised $\Delta V_{on}$ vs position of an L=40 μm channel device to one with L=60 μm, and in both cases the biggest $V_{on}$ shift occurs 5-10 μm away from the source electrode edge. The polycrystalline domains within the channel appear similar in the two cases, with grains elongating from the electrodes towards the centre of the channel, the grains being accordingly longer in the L=60 μm device. While the data in Figure 3a corresponds to a linear scan located roughly halfway within the W=100 μm width of the channel, scans conducted nearer either of the lateral channel edges yield similar results, indicating that all channel sections are equivalent in terms of photoresponse. Notably, neither of the devices of Figure 3a shows any visible response near the grain boundary in the middle of the channel, which the linear scan necessarily crosses in a perpendicular direction. Overall grain boundaries and crystal domain geometry do not appear to play any significant role regarding photoresponse in our experiments. Instead, it is the distance of the probing laser spot from the source electrode that appears to have a critical role. This aspect is further explored by conducting linear illumination scans over interdigitated electrodes with alternating source and drain electrodes: when the OFET is scanned in one direction the laser spot crosses electrodes in the D→S→D→S sequence, whereas in the opposite direction the sequence is S→D→S→D (**Figure 3b-d**). When the spot is moving from a drain to a source electrode, $V_{on}$ shifts as expected upon approaching the source edge. Surprisingly, when instead it moves from a source to a drain electrode, $V_{on}$ remains constant and no photoresponse is observed at all. This contrast is conspicuous in interdigitated electrodes, because in the forward direction $V_{on}$ shifts twice, while in the reverse direction it shifts only once in the middle channel section. This behaviour is observed systematically both in interdigitated and single channel geometries, and in the latter the transfer curve remains fully unchanged even at the end of a full linear scan. When a single channel device is scanned in the reverse direction (S→D) with no change in transfer, it can then be scanned in the forward direction (D→S) in the same or different location within the channel and display a $V_{on}$ shift as expected, so a reverse scan does not affect subsequent photoresponse measurements. Whether the change in scan direction is achieved by inverting the movement direction of the laser, swapping the leads contacting the electrodes, or inverting the voltage applied to the electrodes, is irrelevant and produces the same outcome.

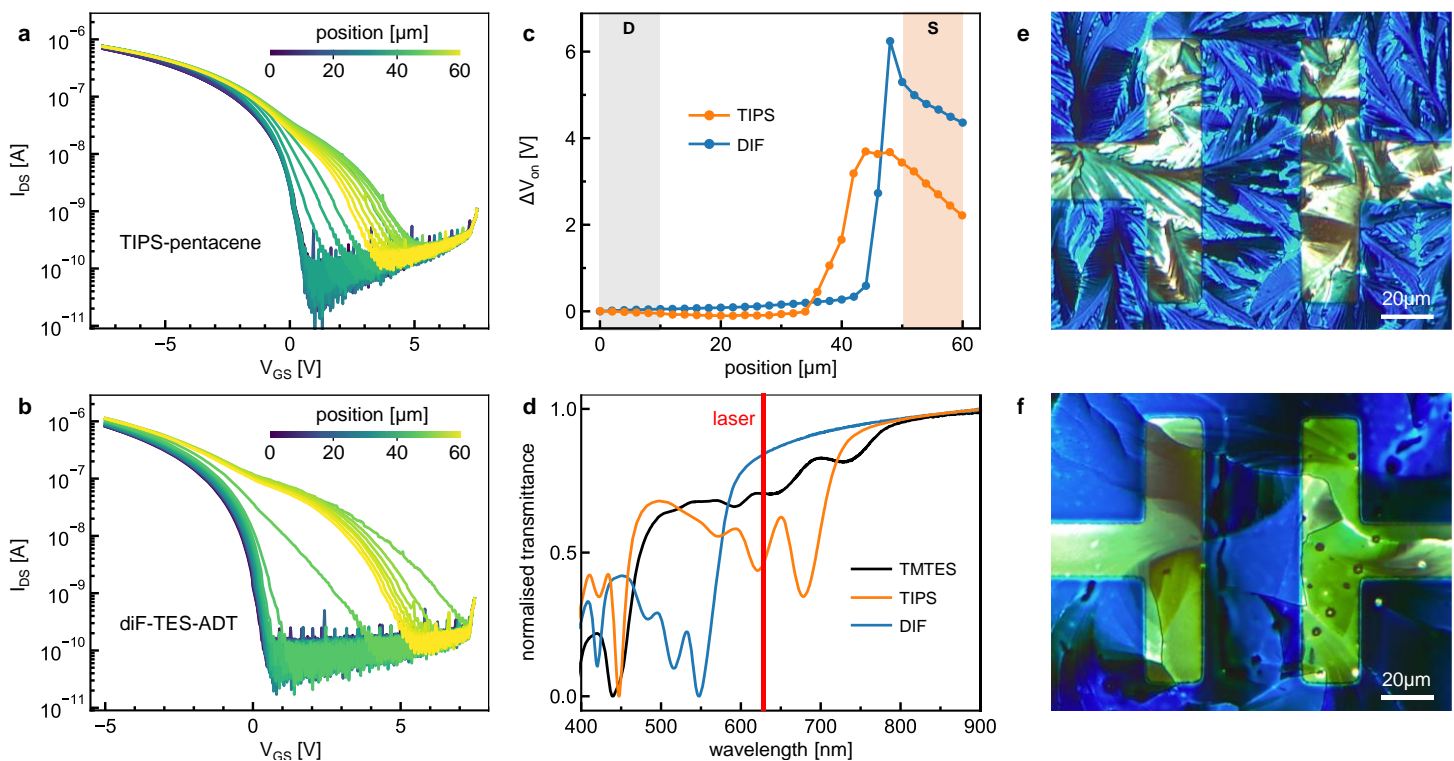

**Figure 4. Photoresponse of TIPS-pentacene and diF-TES-ADT**. **a**, Change in transfer curves along a scan in TIPS-pentacene and **b**, diF-TES-ADT devices. **c**, $\Delta V_{on}$ behaves similarly to TMTES-pentacene for both materials. **d**, Transmittance spectra of each material with marked laser lane wavelength of 635 nm. **e**, Polarised microscopy images of TIPS-pentacene and **f**, diF-TES-ADT OFETs.

Optically probing the channel in random individual points within the channel does not produce any response, instead points must follow a path from drain to source with sufficiently small spacing. We observe that the scan step size is crucial. **Figure 3e** plots the maximum $\Delta V_{on}$ recorded for a set of identical devices scanned with increasing scan step size. The photoresponse drops sharply for step size >6 μm, with steps ≥10 μm producing little to no change in transfer characteristics.

Further, when a channel is scanned at $V_{DS}=0$ V no photoresponse is ever observed and the initial transfer characteristic in the dark is maintained, as is the case if the scan is done by limiting the $V_{GS}$ range such that $V_{GS}>V_{th}$ so the OFET never turns on electrically. A small negative $V_{DS}=-1$ V is already sufficient to trigger a clear shift in $V_{on}$, with the sharp localised increase in $\Delta V_{on}$ clearly emerging for $V_{DS}≤-3$ V (**Figure 3f**). $\Delta V_{on}$ increases further with more negative $V_{DS}$, although the maximum $\Delta V_{on}$ shows variations between devices. Therefore, the photoresponse effect is only observed when charges are flowing along the channel.

In addition to TMTES-pentacene, other benchmark organic semiconductors Bis(triisopropylsilylethynyl) pentacene (TIPS-pentacene) and 2,8-Difluoro-5,11-bis(triethylsilylethynyl) anthradithiophene (diF-TES-ADT) were studied adopting the same OFET geometry and optical probing protocol. The response is qualitatively the same as for TMTES-pentacene (**Figure 4a,b**), with an abrupt shift of $V_{on}$ as the laser spot approaches the source electrode, although a higher optical power of 10 μW is required to achieve a maximum $\Delta V_{on}$ comparable to that of TMTES-pentacene with 1µW power under the same electrical measurement settings (**Figure 4c**). The lower response intensity can be expected for diF-TES-ADT as the laser wavelength falls within a tail of its absorption spectrum (**Figure 4d**), however the absorption of TIPS-pentacene is actually higher than that of TMTES-pentacene in this spectral region. Other factors determining photoresponse intensity could be film thickness and field-effect mobility, which is lowest in TIPS-pentacene and could justify its reduced response. As opposed to TMTES-pentacene, electrodes do not appear to serve as nucleation sites for crystal growth for neither TIPS-pentacene nor diF-TES-ADT (**Figure 4e,f**). In both cases individual crystal domains can be large (>50 µm) and span wide areas of the device including the entire channel, with grain boundaries crossing channel and electrode edges in various locations and directions. No noticeable change in transfer characteristic is observed in the proximity of these grain boundaries upon illumination, supporting the previous observation on TMTES-pentacene that they do not play a significant role in photoresponse.

## Discussion

The key experimental findings can be summarised as follows: i) focussed laser light shifts $V_{on}$ towards positive values, effectively turning on the transistor in $V_{GS}$ ranges where it was previously off, and the effect is persistent after illumination is removed; ii) the response is concentrated at a distance of 5-10 µm from the source electrode, and only occurs if the laser spot reaches that region from the drain electrode with a scan step size <10 µm; iii) $\Delta V_{on}$ scales with illumination dose and $I_{DS}$, and no shift is observed if $I_{DS}=0$ during illumination; iv) maximum $I_{DS}$ and field-effect mobility, as conventionally extracted from the $V<V_{th}$ range for devices in the dark, are largely unaffected.

Localised optical heating can be ruled out as a potential origin of these effects, since they clearly occur at low power levels <1 µW, and the isotropic nature of heating is in contrast with the observed asymmetries between electrodes and the $I_{DS}$ correlation. Interaction of light with the organic semiconductor through interband transitions with creation of excitons is thus more plausible as a general underlying mechanism for the observed behaviour. Once an exciton is formed, it can either recombine without producing any other effect, or dissociate into an electron and hole that lead to further interactions. Photocurrent, defined as the current given by direct collection of the electron and hole of the split exciton in the bulk, cannot play a significant role because the material is p-type, so only hole transport is allowed, and photocurrent should be detected regardless of gate bias. Photoconduction, defined as a change of semiconductor bulk conductance due to increased charge carrier concentration caused by light (equivalent to doping), can also be excluded because no change in conductance is detected with illumination when the transistor is electrically in the off state. Instead, our results are compatible with local trapping of minority carriers (electrons) near the semiconductor-dielectric interface. Exciton dissociation is assisted by the source-drain current at the accumulation layer, and while the generated hole is transported towards the drain together with all other majority carriers, the electron remains locally trapped at the site of illumination. The trapped charge creates a negative potential that works as an additional local gate with a fixed effective voltage. In our system, the radius of this potential is estimated to be around 5 µm, decaying rapidly beyond this distance. Since trapping is long-lived, if another location is optically stimulated nearby an extended region of localised negative potential can be formed within the channel. Our data suggest that once a location sufficiently close to the source electrode is illuminated (within <10 µm), majority carriers can directly inject into this negative potential region and reach the drain, as if travelling along a parallel channel with its own independent gate compared to the rest of the device. The effective voltage of this localised gate is determined by the amount of trapped negative charge, and can reach very high values in terms of actual device bottom gate equivalent (up to 60 V reported above) because the charge is trapped in very close proximity to the accumulation layer, so its effect is amplified. On the other hand, this localised channel is narrow, limited by the width of the induced local potential, so its charge transport capacity is low and dominated by the global channel conductance when the OFET is in the electrically on state for $V_{GS}<V_{th}$. This picture of localised electron trapping is further supported by Kelvin probe force microscopy (KPFM) conducted on the OFET channel after a linear optical probing scan. A region of potential 0.1 V below the average level of the channel is observed along the path followed by the laser spot (**Figure 5a,b**), consistent with an accumulation of negative charges near the surface. The width of this region is around 10-20 µm, compatible with but somewhat larger than the step size required to optically induce a sizeable $\Delta V_{on}$, possibly due to a gradual diffusion of charges with time. The simultaneous topography data collected by the atomic force microscope (AFM) highlights well the crystal domains and grain boundaries, but does not show any notable features along the illumination path (**Figure 5c**).

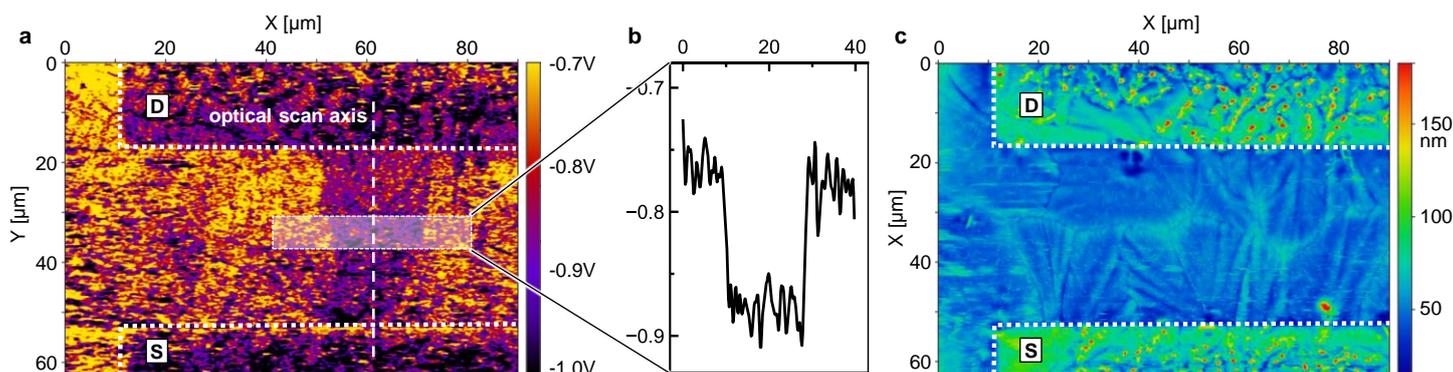

**Figure 5. KPFM of trapped charges**. **a**, KPFM scan of channel area, with negative potential region along scan axis indicating trapped electrons. **b**, Cross section of boxed area in a, averaged along the Y direction. **c**, AFM topography acquired simultaneously with the KPFM scan in a, no topographic features visible along the optical scan axis.

To summarise, in this work we demonstrated controlled trapping of electrons within an OFET channel by localised optical probing. Trapped charges significantly shift the turn-on voltage of the transistor, profoundly changing its electrical characteristics. Trapping is determined by multiple factors including illumination dose and position, source-drain current, and semiconductor material, and these parameters were systematically investigated. These findings unlock the response mechanism underpinning a wide range of organic photodetectors, particularly phototransistors, and pave the way for rational trap engineering in practical devices. Controlled illumination of OFETs could enable tuning of transistor threshold voltage, signal routing in organic circuits, and development of organic optoelectronic memories.

**Methods**.
Materials. 1,4,8,11-tetramethyl-6,13-triethylsilylethynyl pentacene (TMTES-pentacene) and 6,13-Bis(triisopropylsilylethynyl)pentacene (TIPS-pentacene) were purchased from Ossila Ltd. 2,8-Difluoro-5,11-bis(triethylsilylethynyl)anthradithiophene (diF-TES-ADT) was purchased from Lumtec. Polystyrene of molecular weight 10 kg/mol (PS10k) and 280 kg/mol (PS280k), 2,3,4,5,6-pentafluorothiophenol (PFBT) and all organic solvents were purchased from Sigma-Aldrich. All materials were used as-is without further purification.
Ink formulations. TMTES-pentacene and PS280k powders were separately dissolved in anhydrous chlorobenzene in 2wt% concentration by heating at 105°C for 2 hours. They were then mixed in a 2:1 volume ratio of TMTES-pentacene:PS280k, and left at 60°C overnight until use. TIPS-pentacene and diF-TES-ADT ink formulations followed the same protocol, but using PS10k and a 4:1 semiconductor:PS mixing ratio.
Substrate preparation. Samples were fabricated on p-doped Si wafer substrates from Si-Mat, with resistivity <0.02 Ωcm and 200 nm thick thermal oxide coating. Electrodes were defined by photolithography using positive resist exposed with a MicroWriter ML3 system by Durham Magneto Optics Ltd. After resist development, 5 nm Cr + 50 nm Au layers were deposited by thermal evaporation, followed by lift-off in acetone and isopropanol. Substrates were stored in a nitrogen glovebox until semiconductor coating. The electrode pattern yields 3 OFET devices per sample, each with its separate electrodes and contact pads. Devices have a bottom gate, bottom contact geometry, where source and drain electrodes sit on top of the planar gate dielectric ($SiO_2$) and are covered by the semiconductor layer. The bulk Si is used as common gate for all devices within one sample.

<u>Semiconductor deposition</u>. Prior to coating, substrates were treated with UV-ozone for 25 minutes, then immersed in a 15 mM solution of PFBT in isopropanol for 15 minutes to form a self-assembled monolayer on the electrodes. The semiconductor was applied by bar-assisted meniscus shearing technique with the prepared inks as previously reported,[24,27] using a substrate temperature of 105°C and a bar speed of 28 mm/s. The coating was manually patterned to define a 1 mm² patch on each device centred on the OFET channel region, by wiping the sample surface after coating with a cleanroom wipe lightly soaked in acetone.

<u>Electrical and optical measurements</u>. Samples are characterised in ambient conditions using a Keithley 2612B source measure unit with a set of probes integrated into a custom microscopy setup. The $SiO_2$ dielectric is scraped near the sample edge to access the bottom Si with a sharp probe for OFET gating. Sample and probes are mounted on a pair of CONEX-MFACC stages (Newport Corporation) for XY position scanning. Laser light is coupled into the microscope from a Thorlabs MCLS1 source and focussed with an Olympus MPLFLN40X objective with 40× magnification and numerical aperture of 0.75.

<u>Spectrophotometry</u>. Transmittance spectra for all materials were acquired with a Jasco V-770 UV-Visible/NIR spectrophotometer, using plain glass microscope coverslips as substrates and following the protocol detailed above for semiconductor deposition.

<u>Imaging</u>. Polarisation imaging was done ex situ using an Olympus BX51 microscope equipped with polariser and analyser.

<u>AFM and KPFM</u>. Topography AFM and KPFM images were acquired ex situ with a Park System NX10 atomic force microscope, and data analysed with the software Gwyddion.

**Acknowledgements**.

D. K. acknowledges funding from Horizon Europe under the Marie Skłodowska-Curie project OPTOCHARGE (grant agreement No. 101066319), and Beatriu de Pinós programme by Agència de Gestió d'Ajuts Universitaris i de Recerca (grant agreement 2023-BP-00236). This work was also funded by MCIN/AEI/10.13039/501100011033/ERDF,UE with projects SENSATION PID2022-141393OB-I00 and PDC2022-133750-I00, and through the "Severo Ochoa" Programme for Centers of Excellence in R&D (CEX2023-001263-S). The authors also thank Generalitat de Catalunya (2021-SGR-00443).